\documentclass[aps,prl,preprint,tightenlines,superscriptaddress,showpacs,byrevtex]{revtex4}
\usepackage{graphicx} 
\usepackage{dcolumn}  

\graphicspath{{ps}}

\newcommand{\bbar}{\overline{B}{}^{\,0}}

\newcommand{\ra}{\!\rightarrow\!}

\begin{document}


\preprint{\vbox{ \hbox{   }
                 \hbox{BELLE-CONF-0555}
                 \hbox{LP2005-190}
		 \hbox{EPS05-530} 
}}
\title{ \boldmath\quad\\[0.5cm]  Improved measurement of $\bbar\to D_s^-D^+$ and \\
search for $\bbar\to D_s^+D_s^-$ at Belle\unboldmath}

\affiliation{Aomori University, Aomori}
\affiliation{Budker Institute of Nuclear Physics, Novosibirsk}
\affiliation{Chiba University, Chiba}
\affiliation{Chonnam National University, Kwangju}
\affiliation{University of Cincinnati, Cincinnati, Ohio 45221}
\affiliation{University of Frankfurt, Frankfurt}
\affiliation{Gyeongsang National University, Chinju}
\affiliation{University of Hawaii, Honolulu, Hawaii 96822}
\affiliation{High Energy Accelerator Research Organization (KEK), Tsukuba}
\affiliation{Hiroshima Institute of Technology, Hiroshima}
\affiliation{Institute of High Energy Physics, Chinese Academy of Sciences, Beijing}
\affiliation{Institute of High Energy Physics, Vienna}
\affiliation{Institute for Theoretical and Experimental Physics, Moscow}
\affiliation{J. Stefan Institute, Ljubljana}
\affiliation{Kanagawa University, Yokohama}
\affiliation{Korea University, Seoul}
\affiliation{Kyoto University, Kyoto}
\affiliation{Kyungpook National University, Taegu}
\affiliation{Swiss Federal Institute of Technology of Lausanne, EPFL, Lausanne}
\affiliation{University of Ljubljana, Ljubljana}
\affiliation{University of Maribor, Maribor}
\affiliation{University of Melbourne, Victoria}
\affiliation{Nagoya University, Nagoya}
\affiliation{Nara Women's University, Nara}
\affiliation{National Central University, Chung-li}
\affiliation{National Kaohsiung Normal University, Kaohsiung}
\affiliation{National United University, Miao Li}
\affiliation{Department of Physics, National Taiwan University, Taipei}
\affiliation{H. Niewodniczanski Institute of Nuclear Physics, Krakow}
\affiliation{Nippon Dental University, Niigata}
\affiliation{Niigata University, Niigata}
\affiliation{Nova Gorica Polytechnic, Nova Gorica}
\affiliation{Osaka City University, Osaka}
\affiliation{Osaka University, Osaka}
\affiliation{Panjab University, Chandigarh}
\affiliation{Peking University, Beijing}
\affiliation{Princeton University, Princeton, New Jersey 08544}
\affiliation{RIKEN BNL Research Center, Upton, New York 11973}
\affiliation{Saga University, Saga}
\affiliation{University of Science and Technology of China, Hefei}
\affiliation{Seoul National University, Seoul}
\affiliation{Shinshu University, Nagano}
\affiliation{Sungkyunkwan University, Suwon}
\affiliation{University of Sydney, Sydney NSW}
\affiliation{Tata Institute of Fundamental Research, Bombay}
\affiliation{Toho University, Funabashi}
\affiliation{Tohoku Gakuin University, Tagajo}
\affiliation{Tohoku University, Sendai}
\affiliation{Department of Physics, University of Tokyo, Tokyo}
\affiliation{Tokyo Institute of Technology, Tokyo}
\affiliation{Tokyo Metropolitan University, Tokyo}
\affiliation{Tokyo University of Agriculture and Technology, Tokyo}
\affiliation{Toyama National College of Maritime Technology, Toyama}
\affiliation{University of Tsukuba, Tsukuba}
\affiliation{Utkal University, Bhubaneswer}
\affiliation{Virginia Polytechnic Institute and State University, Blacksburg, Virginia 24061}
\affiliation{Yonsei University, Seoul}
  \author{K.~Abe}\affiliation{High Energy Accelerator Research Organization (KEK), Tsukuba} 
  \author{K.~Abe}\affiliation{Tohoku Gakuin University, Tagajo} 
  \author{I.~Adachi}\affiliation{High Energy Accelerator Research Organization (KEK), Tsukuba} 
  \author{H.~Aihara}\affiliation{Department of Physics, University of Tokyo, Tokyo} 
  \author{K.~Aoki}\affiliation{Nagoya University, Nagoya} 
  \author{K.~Arinstein}\affiliation{Budker Institute of Nuclear Physics, Novosibirsk} 
  \author{Y.~Asano}\affiliation{University of Tsukuba, Tsukuba} 
  \author{T.~Aso}\affiliation{Toyama National College of Maritime Technology, Toyama} 
  \author{V.~Aulchenko}\affiliation{Budker Institute of Nuclear Physics, Novosibirsk} 
  \author{T.~Aushev}\affiliation{Institute for Theoretical and Experimental Physics, Moscow} 
% ...
\collaboration{The Belle Collaboration}


\begin{abstract}
We reconstruct $\bbar\to D_s^-D^+$ decays using a sample of
$275\times 10^6$~$B\bar{B}$ pairs recorded by the Belle experiment, and
we measure the branching fraction to be $[7.42\pm 0.23({\rm stat.})\pm 1.36({\rm syst.})]\times 10^{-3}$.
We also search for the related decay $\bbar\to D_s^+D_s^-$.
We observe no statistically significant signal and set
an upper limit on the branching fraction of $2.0\times 10^{-4}$ at 90\% C.L.
\end{abstract}

\pacs{13.25.Hw, 14.40.Nd}

\maketitle

\tighten

{\renewcommand{\thefootnote}{\fnsymbol{footnote}}}
\setcounter{footnote}{0}

\subsection{Introduction}
Several decay modes of $B$ mesons with $D^+_s$ in the
final state have been measured at the $B$ factories.
The amplitudes governing these decays are interesting
because neither constituent flavor of the $D^+_s$
is present in the initial state.
For example, the recently-observed decays
$\bbar\ra D^+_s K^-$~\cite{Dspi} and
$\bbar\ra D^+_{sJ}(2317) K^-$~\cite{dsjk}, with branching fractions in 
the range $10^{-5}-10^{-4}$, can proceed via a $b\bar{d}\ra c\bar{u}$ $W$-exchange diagram. 
Here we study the related decays
$\bbar\ra D^+_s D^-_s$ and $\bbar\ra D^-_s D^+$.
The former proceeds via Cabibbo-suppressed $W$-exchange
and has not yet been observed;
a recent theoretical calculation predicts a branching fraction
of $\sim$\,$2.5\times 10^{-4}$~\cite{Fajfer}. The latter
proceeds via a Cabibbo-favored tree diagram; the ratio of
its branching fraction to that for $\bbar\ra D^+_s\pi^-$
can be used to determine the ratio of CKM matrix elements
$|V^{}_{ub}/V^{}_{cb}|$~\cite{CKM} assuming factorization~\cite{Kim}. However, the
PDG value~\cite{PDG} for $B(\bbar\ra D^+_s D^-)$ has
a large uncertainty (38\%), which limits the usefulness
of this method at present.

In this paper we report an improved measurement of $\bbar\to D_s^-D^+$ decays and a search for $\bbar\to
D_s^+D_s^-$ decays with the Belle detector~\cite{Belle_det} at the
KEKB asymmetric energy $e^+e^-$ collider~\cite{KEKB_acc}. The results
are based on a $253$ fb$^{-1}$ data sample collected at the center-of-mass (CM)
energy of the $\Upsilon(4S)$ resonance. 
To study backgrounds, we used a Monte Carlo (MC)
simulated sample of $\Upsilon(4S)\to B\bar{B}$ events and $e^+e^-\to q\overline{q}$ ($q=u$, $d$, $s$ and $c$ quarks)
continuum events; the sample size was about twice that of
the data. To evaluate the reconstruction efficiency,
we used a large MC sample of several thousand signal events.

The Belle detector is a large-solid-angle magnetic spectrometer that consists of a multi-layer silicon
vertex detector (SVD), a 50-layer central drift chamber (CDC), an array of aerogel threshold
\v{C}erenkov counters (ACC), a barrel-like arrangement of time-of-flight scintillation counters
(TOF), and an electromagnetic calorimeter (ECL) comprised of CsI(Tl) crystals located
inside a superconducting solenoid coil that provides a $1.5$ T magnetic field. An iron flux-return
located outside of the coil is instrumented to detect $K^0_L$ mesons and to identify muons (KLM).
The detector is described in detail elsewhere~\cite{Belle_det}. Two different inner detector
configurations were used. For the first 152 million $B\bar{B}$ pairs, a $2.0$ cm radius
beampipe and a 3-layer silicon vertex detector were used; for the latter 123 million $B\bar{B}$ pairs,
a $1.5$ cm radius beampipe, a 4-layer silicon detector and a small-cell inner drift chamber were
used~\cite{Belle2}.

Charged tracks were selected with loose requirements on the impact parameter
relative to the interaction point (IP). We also required the transverse momentum of the tracks
to be greater than $0.2$ GeV$/c$ in order to reduce low momentum combinatorial background. 
Based on the optimization of the product of the efficiency and sample purity as evaluated by MC, we 
impose the following requirements.  
For charged particle identification (PID) we combined information from the CDC, TOF and ACC counters 
into a likelihood ratio ${\cal{L}}(K^\pm)/({\cal{L}}(K^\pm)+{\cal{L}}(\pi^\pm))$~\cite{Belle_ident}, which we 
required to be larger(smaller) than 0.5 for charged kaon(pion) candidates. This requirement selected 
kaons and pions with average efficiencies of 92\% and 96\%, respectively. 
Neutral kaons ($K^0_S$) were reconstructed using pairs of oppositely-charged tracks that have an invariant mass
within 30 MeV$/c^2$ of the nominal $K^0$ mass, and a displaced vertex from the IP.
We used the $D_s^-\rightarrow\phi\pi^-$, $K^{*0}K^-$, and $K^0_SK^-$ modes to reconstruct $D_s^-$ mesons and
$D^+\rightarrow K^+K^-\pi^+$, $K^- \pi^+\pi^+$, and $K^0_S\pi^+$ for the $D^+$ mesons~\cite{foot_1}.
Combinations of oppositely-charged kaons with 
$|M_\phi-M_{K^+K^-}|<20$~MeV/c$^2$ and of oppositely-charged kaons and
pions with $|M_{K^{\ast 0}}-M_{K^+\pi^-}|<85$~MeV/c$^2$
were retained as $\phi$ and $K^{\ast 0}$ candidates, respectively,
where $M_\phi$ and $M_{K^{\ast 0}}$ are the nominal masses of the two mesons~\cite{PDG}. 
These combinations, as well as $K_S^0$'s, were then combined with a
$K^\pm$ or $\pi^\pm$ to form $D_s^\pm$ mesons. 
All combinations with invariant masses within a $4~\sigma$($4.5~\sigma$)
interval around the  nominal $D_s^\pm$($D^\pm$) mass were considered for further
analysis. The values of $\sigma$ were determined from invariant mass distributions 
of MC $D_s^\pm$($D^\pm$) decays and were in the range $3.6-4.2$~MeV/c$^2$.

Pairs of $D_s^-$ and $D^+_{(s)}$ meson candidates were combined to form $\bbar$ meson candidates. These
were identified by their CM energy difference,
$\Delta E=E_B^{\rm CM}-E_{\rm beam}^{\rm CM}$,
and the beam constrained mass,
$M_{\rm bc}=\sqrt{(E_{\rm beam}^{\rm CM})^2-(p_B^{\rm CM})^2}$, 
where $E_{\rm beam}^{\rm CM} = \sqrt{s}/2$ is the CM beam energy and $E_B^{\rm CM}$ and $p_B^{\rm CM}$ are
the reconstructed energy and momentum of the $B$ meson candidate in the CM frame. We retained events having
$M_{\rm bc}>5.2$ GeV$/c^2$ and $|\Delta E| < 0.2$ GeV. The signal region was defined as 
$5.272$ GeV$/c^2$ $\leq$ $M_{\rm bc}$ $\leq$ $5.285$ GeV$/c^2$ and $|\Delta E|\leq$ $0.025$ GeV.

To suppress the large combinatorial background dominated by the two-jet-like $e^+e^-\to q\bar{q}$
continuum process, variables characterizing the event topology were used. We
required the ratio of the second to zeroth Fox-Wolfram moments~\cite{fox-wolfram} $R_2<0.3$ and the thrust value 
of the event $T<0.8$.
Simulation showed that this selection retained more than 95\% of $B\bar{B}$ events and rejected about 55\% of
$c\bar{c}$ events and 65\% of $u\bar{u}$, $d\bar{d}$ and $s\bar{s}$ events.

\subsection{\boldmath $\bbar\to D_s^-D^+$ decays\unboldmath}
\begin{figure}[t]
\begin{center}
\includegraphics[width=0.49\textwidth]{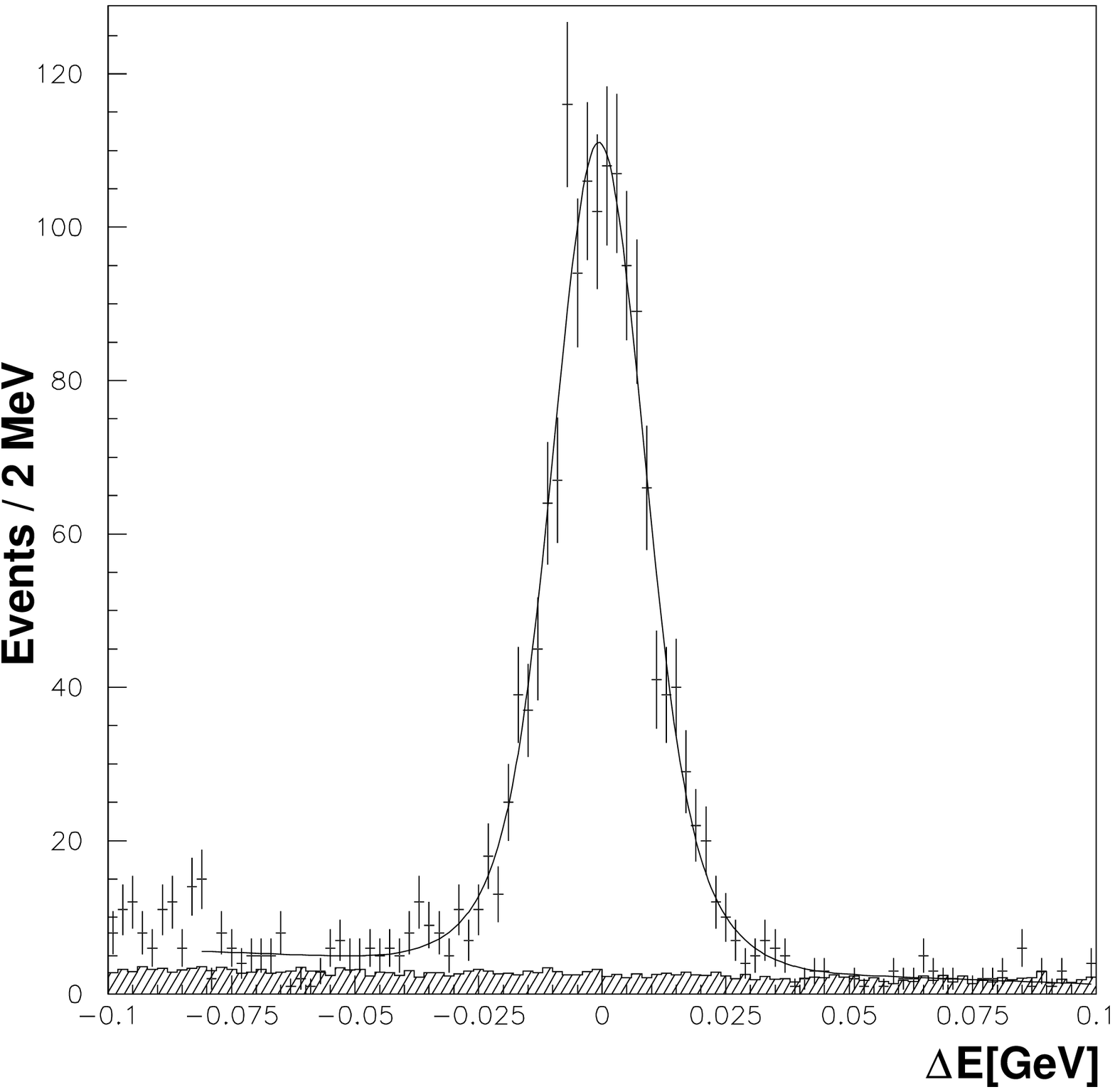}
\includegraphics[width=0.49\textwidth]{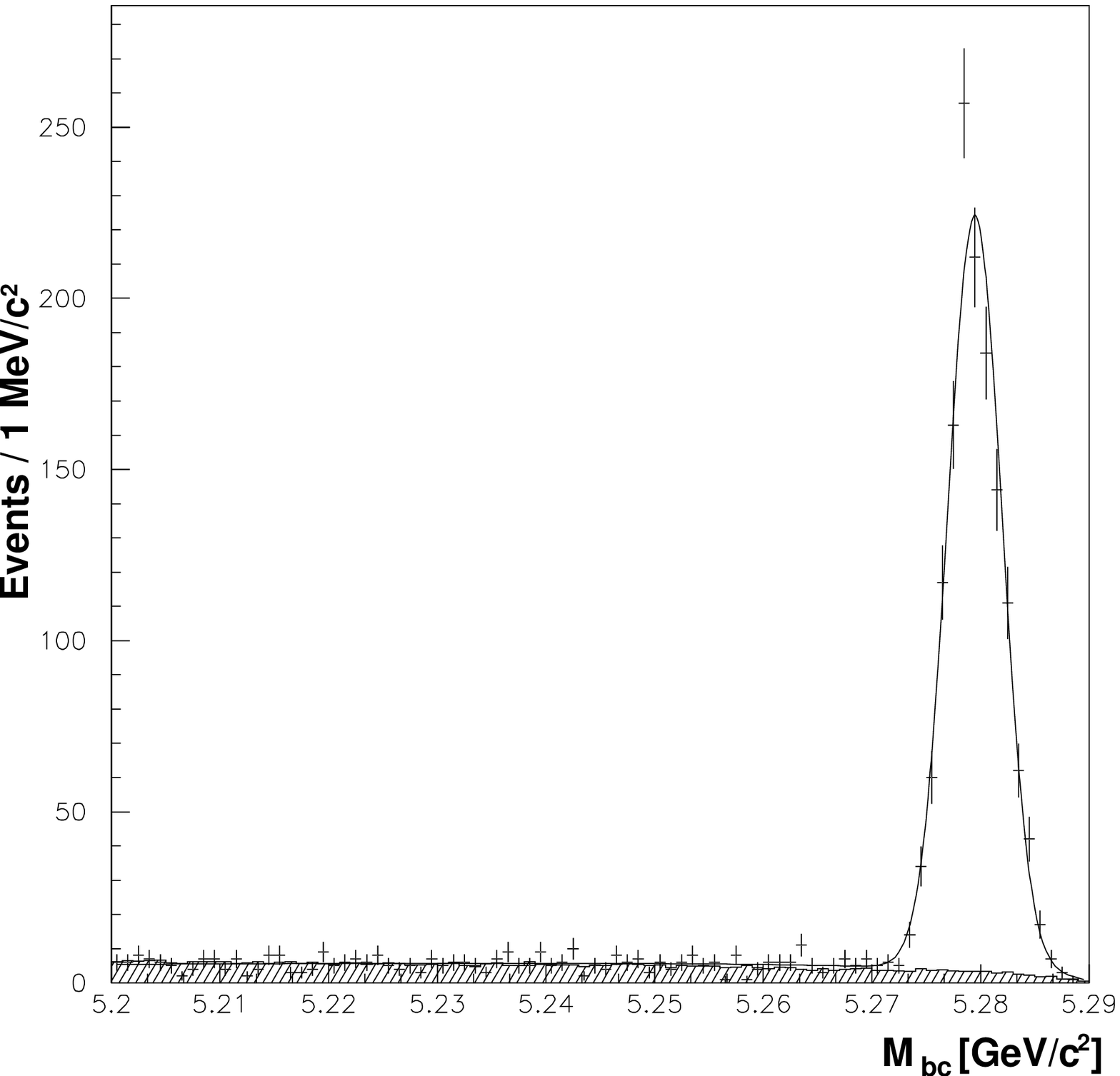}
\end{center}
\caption{$\Delta E$ (left) and $M_{\rm bc}$ (right) distributions for reconstructed $\bbar\to D_s^-D^+$ 
events in the $M_{\rm bc}$ or $\Delta E$ signal region, respectively. The curve shows the results of 
the fit. The normalized distributions for the events in the sideband of $D_s$ and $D$ meson masses are shown 
as the hatched histograms.} 
\label{fig_8}
\end{figure}

The $\Delta E$ and $M_{\rm bc}$ distributions for $\bbar\to D_s^-D^+$ decays obtained after applying all selection criteria
described above are shown in Fig.~\ref{fig_8}. 

The beam-constrained mass distribution is well-described using a 
Gaussian function for the signal
and an empirically parameterized background shape introduced by the ARGUS Collaboration~\cite{Argus}. 
The energy difference is
described using two Gaussians with the same mean for the signal and a linear function for 
the background. The normalizations, positions and widths of the signal are free parameters of the 
binned likelihood fit. The solid lines in Fig.~\ref{fig_8} show the
result of fits in the interval $-0.08$ GeV $<$ $\Delta E$ $<$ $0.10$ GeV and $5.20$ GeV$/c^2$ $<$ $M_{\rm bc}$ $<$ $5.29$ GeV$/c^2$. 
The fit to the $\Delta E$ distribution for events in the
$M_{\rm bc}$ signal region gives a signal yield of $N_{\rm data}=1372\pm 42$ events. The width of the narrower
Gaussian that contains about $66\%$ of the signal is $8.6\pm 0.9$ MeV.
From the fit to the $M_{\rm bc}$ distribution for events in 
the $\Delta E$ signal region we find $1381\pm 45$ signal events. 
Due to the simpler 
parameterization of the energy difference distribution and possible peaking background
in the $M_{\rm bc}$ distribution, we choose to evaluate all the results from the
fits to the $\Delta E$ distribution for events within the $M_{\rm bc}$ signal region.
While the level of
the combinatorial background is found to be low, we used events in the sidebands of the $D_s$ and $D$ meson 
invariant mass distributions to check for possible remaining peaking background as described later.  

The efficiency of the selection criteria depends weakly on the reconstructed decay 
channel of the charmed mesons. Efficiencies as obtained 
from the simulated event sample for different combinations of $D$ and $D_s$ meson decay channels are 
used for calculation of the
overall efficiency. 
The efficiency includes intermediate branching fractions:  
$\epsilon (D_sD)=\sum_{i,j=1,3}\epsilon_{i,j}{\cal B}(D_{si}){\cal B}(D_j)=(6.31\pm 0.88)\times 10^{-4}$.
Here, $\epsilon_{i,j}$ represents the efficiency for reconstructing the event if the $D_s$ meson decays through 
the $i$-th mode and the $D$ meson through the $j$-th mode. 
The largest part of the uncertainty ($\pm 13.9$\%) arises due to the uncertainties in the intermediate branching fractions~\cite{PDG,DsBr}. 
There is also a small contribution due to limited MC statistics. 
Alternatively, one can express the efficiency in terms of that for
${\cal B}(D_s^+\to\phi\pi^+)$, 
relative to which the branching fractions for $D_s^-\to K^{\ast 0}K^-$ and $D_s^-\to K_S^0K^-$ are well-measured~\cite{PDG}. In this case we obtain 
$\epsilon (D_sD)=(3.18\pm 0.25)\times 10^{-2} \cdot {\cal B}(D_s^+\to\phi\pi^+){\cal B}(\phi\to K^+K^-)$.

The signal peak includes non-resonant $D_s^-\to K^+K^-\pi^-$ decays 
resulting in the same final state. By fitting the $\Delta E$ distributions of a simulated sample we evaluate the fraction of
such decays with respect to reconstructed $\bbar\to D_s^-D^+$ decays to be 
$r_{KK\pi}=(3.46\pm 0.74)\%$. For the final evaluation of the systematic error we include the uncertainty 
on the branching fraction for the non-resonant $D_s^-\to K^+K^-\pi^-$ decays~\cite{PDG}. This increases the error on
$r_{KK\pi}$, which becomes $(3.46\pm 1.70)\%$. Due to their small contribution to the signal, the nonresonant component is not included in the
branching fraction calculation.

\begin{figure}[t]
\begin{center}
\includegraphics[width=0.49\textwidth]{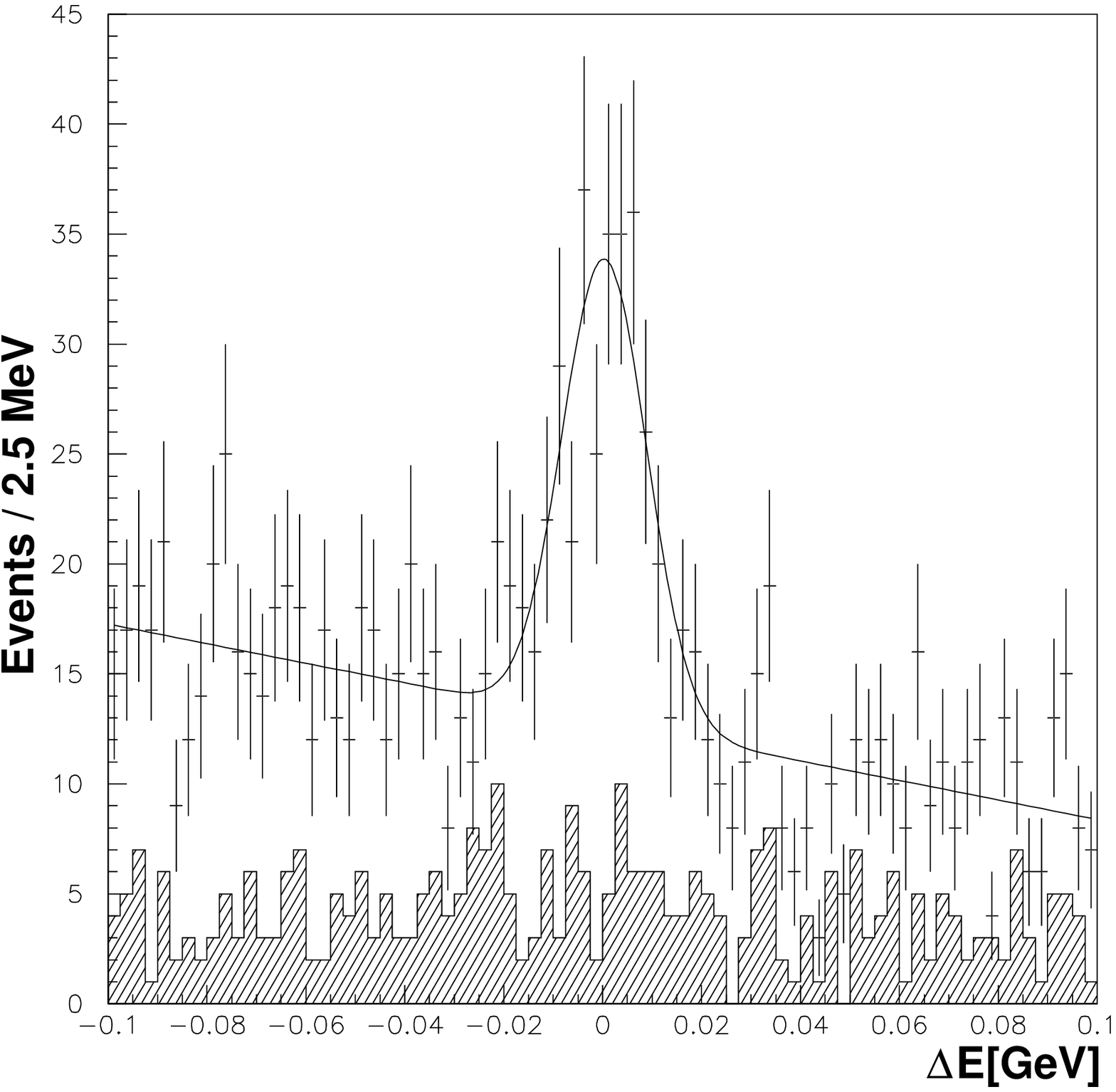}
\includegraphics[width=0.49\textwidth]{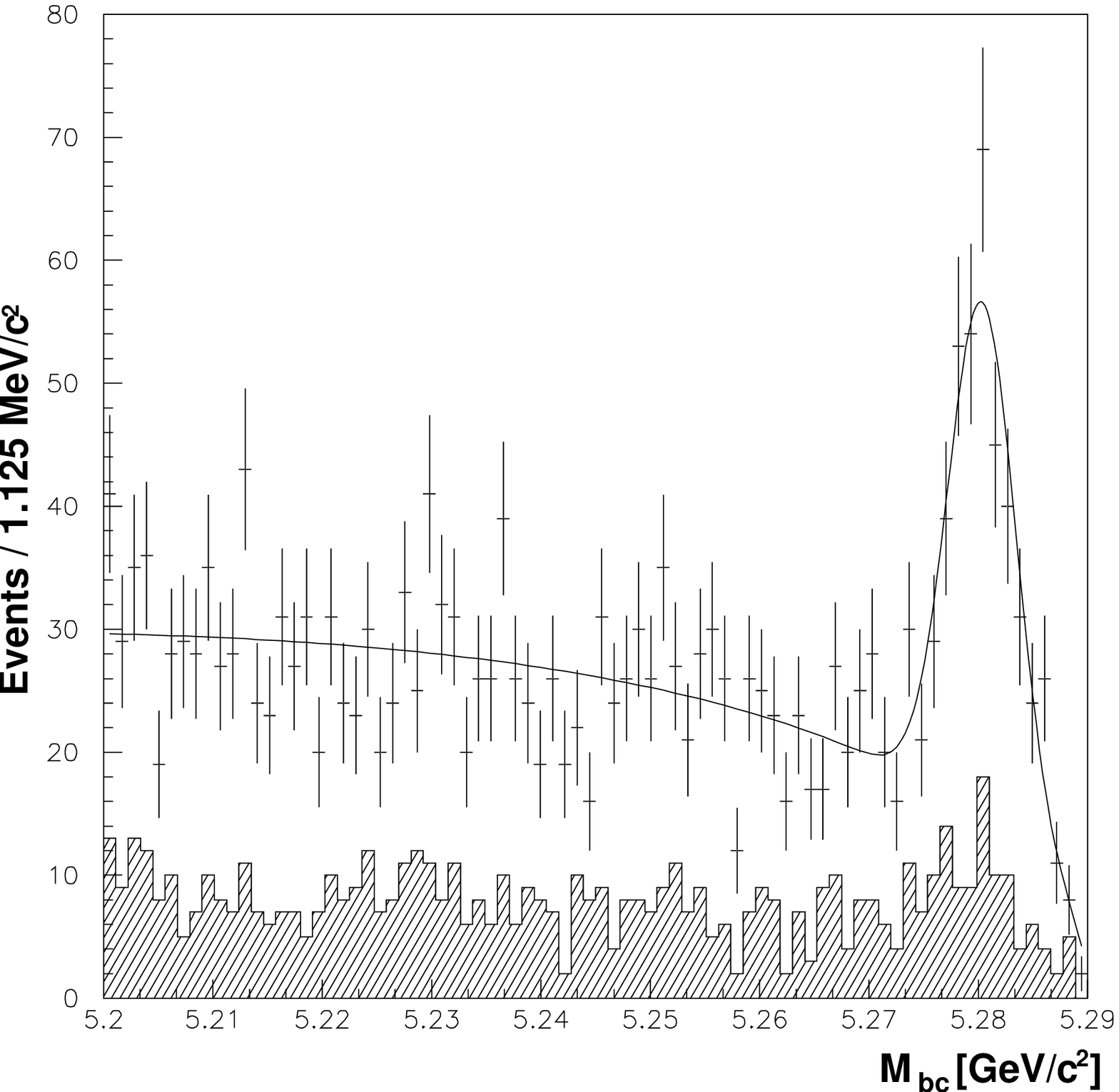}
\end{center}
\caption{$\Delta E$ (left) and $M_{\rm bc}$ (right) distribution for the $D_s$ sideband.
The hatched histogram is the contribution of events reconstructed in the $D_s^-\to
\phi\pi^-$ and $D_s^-\to K^0_sK^-$ decay modes where no significant peaks
are observed. 
The full line is the result of the fit used to determine the 
$\bbar\to D^+ K^{\ast 0} K^-$ contribution.}
\label{fig_11}
\end{figure}

The $\Delta E$ and $M_{\rm bc}$ distributions obtained using events in the sidebands of 
both the $D$ and $D_s$ mesons are in agreement with the observed background under the $\bbar\to
D_s^-D^+$ signal. No peaking is observed for events in the $D$ mass sideband;
however, peaking is observed in the $D_s$ mass sideband as shown in Fig.~\ref{fig_11}.
A peaking component in the $D_s^-\to
K^{\ast 0}K^-$ decay mode is due to the three-body decay 
$\bbar\to D^+ K^{\ast 0} K^-$, previously reported by Belle~\cite{DKstK}. 
Since the $K^{\ast 0} K^-$ invariant mass
in these decays also populates the region under the $D_s^-$ peak, we must subtract this contribution from
the signal. By fitting the $\Delta E$ distribution of the $D_s$ sideband and normalizing the
yield to the area under the $D_s$ peak, we determine the fraction
of such three-body decays within the $\bbar\to D_s^- D^+$ signal peak 
to be $r_{DK^\ast K}=(2.74\pm 0.57)\%$. Since the method assumes approximately equal contributions of 
$\bbar\to D^+ K^{\ast 0} K^-$ decays in the $D_s$ sidebands and in the signal region, 
we assign an additional systematic error to $r_{DK^\ast K}$. From the $K^{\ast 0} K^-$ invariant mass
distribution given in~\cite{DKstK}, we estimate a possible relative
change in the yield over the relevant $M(K^{\ast 0} K^-)$ interval ($1.77$~GeV$/c^2$~$\le M(K^{\ast 0} K^-)\le 2.17$~GeV$/c^2$) 
to be $\pm 40\%$. With the inclusion of this additional uncertainty, the
fraction $r_{DK^\ast K}$ is $(2.74\pm 1.24)\%$. The signal peak in Fig.~\ref{fig_11} also includes a small fraction
of $D_s$ signal that populates the $D_s$ sideband. This contribution was evaluated using the simulated
sample and subtracted from the fitted number of events in the peak.

Taking into account the contributions from non-resonant 
$D_s^-\to K^+K^-\pi^-$ decays (evaluated using
MC simulated events) and from $\bbar\to D^+ K^{\ast 0} K^-$ decays, we determine the
number of signal $\bbar\to D_s^- D^+$ in the sample to be~\cite{foot_2}
$N_{D_sD}=\bigl(1-r_{KK\pi}-r_{DK^\ast K}\bigr)N_{\rm data}=1287\pm 40$,
where the error is statistical only. 

\subsection{\boldmath$\bbar\to D_s^+D_s^-$ decays \unboldmath}

The selection optimized for the reconstruction of $D_s^-$ decays in the $\bbar\to D_s^-D^+$ 
events was also used to search for $\bbar\to D_s^+D_s^-$ decays. In this case two $D_s$ 
meson candidates were searched for in each event using the decay modes listed above.
The $M_{\rm bc}$ and $\Delta E$ variables are then calculated.  

An additional source of background in this decay channel is a cross-feed from $\bbar\to
D_s^-D^+$ decays, where the $D^+$ decays into a $K^-\pi^+\pi^+$ or
$\bar{K}^0\pi^+$ final state, and one of the pions is
misidentified as a kaon. Although these events are shifted from zero in the $\Delta E$
distribution, they make observation of a small possible signal difficult. 
In order to reduce this contribution, we calculated 
for each candidate $D_s^\pm\to K^+K^-\pi^\pm$ or $K^0_SK^{\pm}$ decay 
the invariant mass of the final states particles with the $\pi^\pm$ mass
substituted for the $K^\pm$ mass. If the invariant mass was within $20$~MeV$/c^2$ of the nominal
$D^\pm$ mass the event was rejected. 
The MC efficiencies for the reconstruction of $\bbar\to D_s^+D_s^-$ events for different
combinations of $D_s^\pm$ decay modes are determined and used for calculation of the overall efficiency 
$\epsilon (D_sD_s)=\sum_{i,j=1,3}\epsilon_{i,j}{\cal B}(D_{si}){\cal B}(D_{sj})=(1.63\pm 0.39)\times 10^{-4}$.
Again, the main uncertainty ($\pm 23.1\%$) arises from the uncertainty in the intermediate branching fractions~\cite{PDG,DsBr}. 
The same efficiency expressed in terms of ${\cal B}(D_s^+\to\phi\pi^+)$ reads
$\epsilon (D_sD_s)=(0.383\pm 0.040)\cdot ({\cal B}(D_s^+\to\phi\pi^+){\cal B}(\phi\to K^+K^-))^2$.

\begin{figure}[t]
\begin{center}
\includegraphics[width=0.49\textwidth]{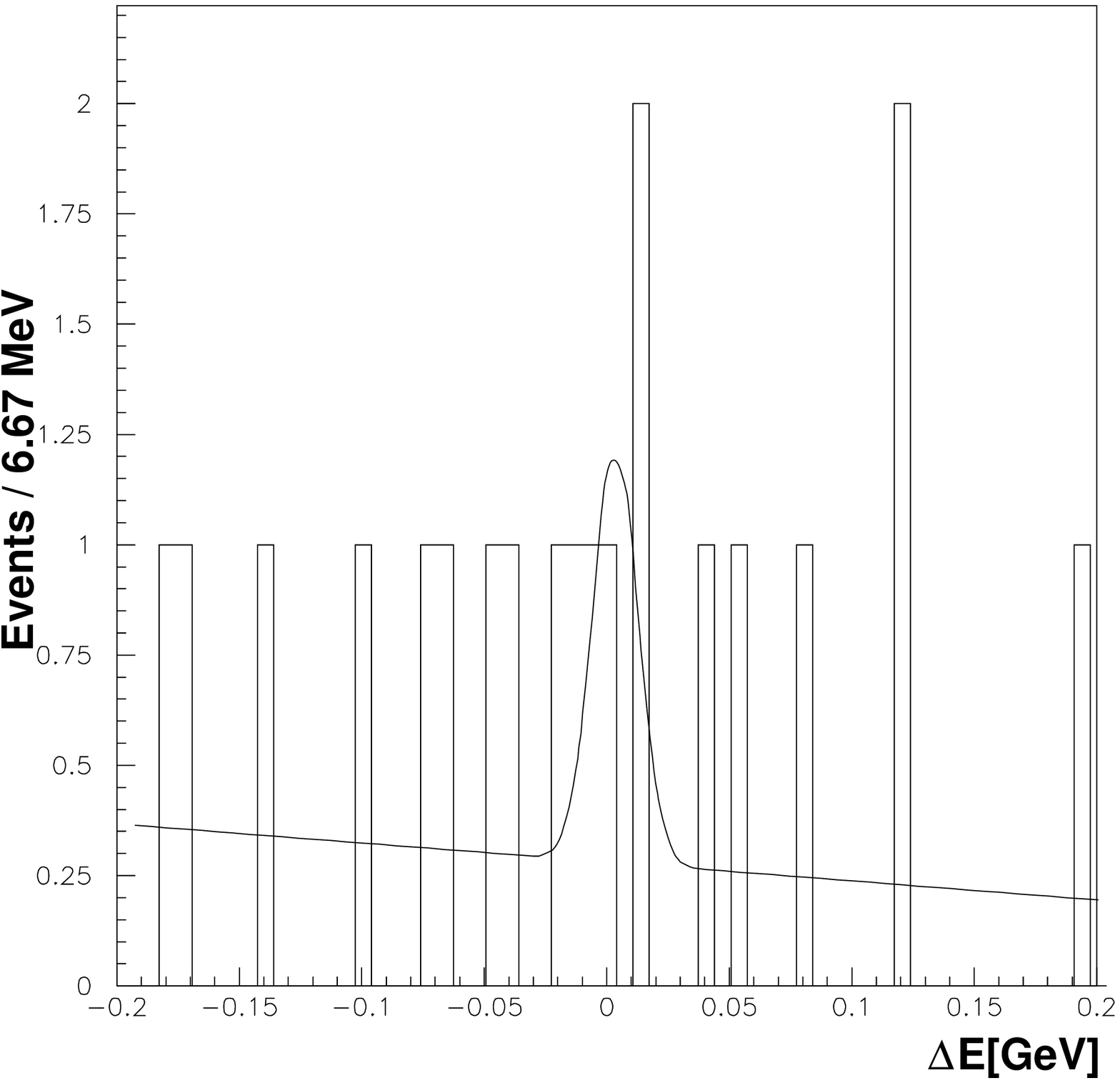}
\includegraphics[width=0.49\textwidth]{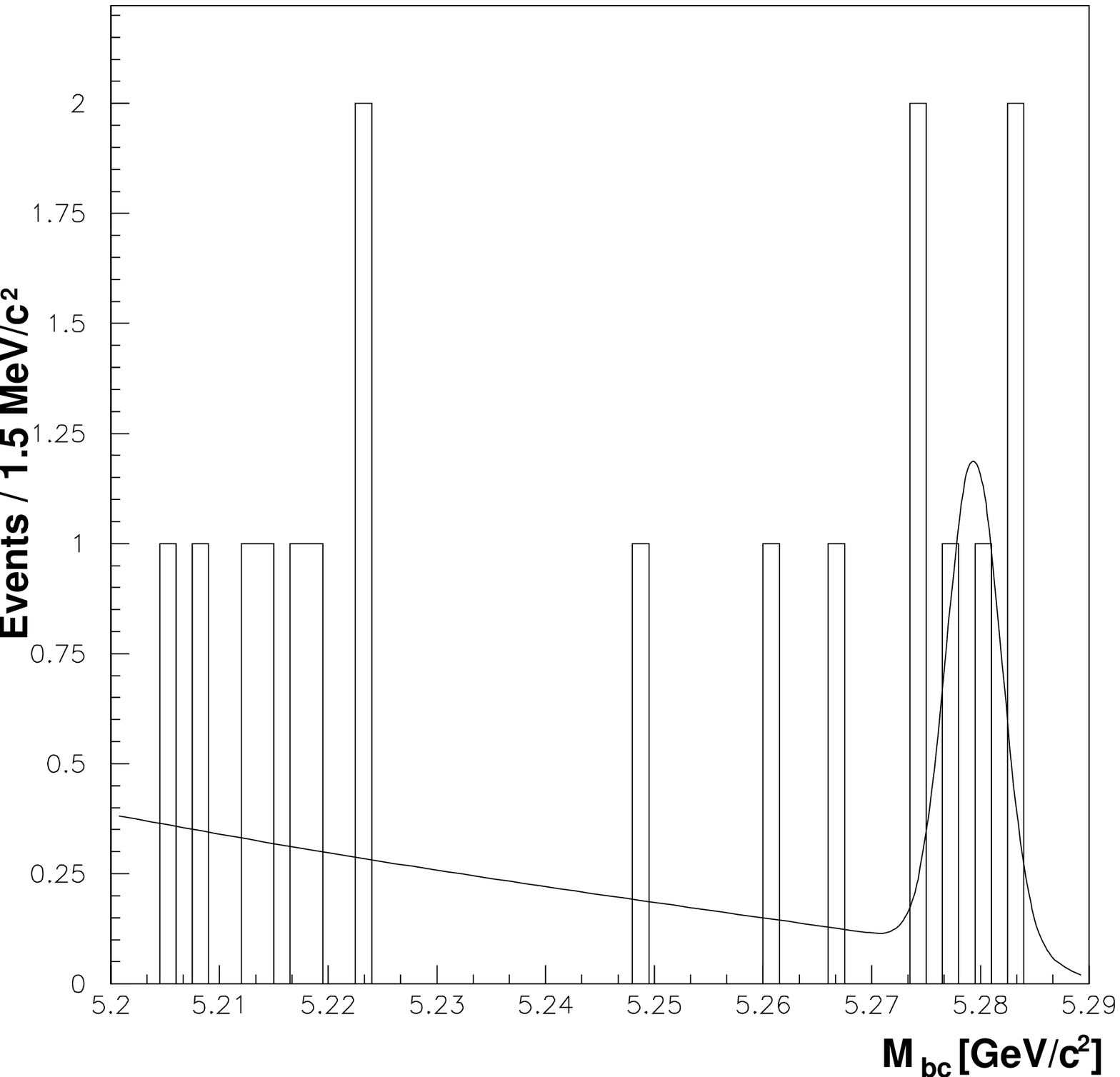}
\end{center}
\caption{$\Delta E$ (left) and $M_{\rm bc}$ (right) distributions for the 
$\bbar\to D_s^+D_s^-$ decay mode.}
\label{fig_12}
\end{figure}

The distributions of $\Delta E$ and $M_{\rm bc}$ for $\bbar\to D_s^+D_s^-$ candidates are shown 
in Fig.~\ref{fig_12}. No significant signal is observed. The distributions are fitted using 
a parameterization of the $D_s^-D^+$ mode. 
However, the signal $\Delta E$ distribution is described 
by a single Gaussian and the position and the width of the signal are fixed to the values obtained 
from the MC and rescaled by corresponding factors evaluated for the $D_s^-D^+$ mode. 
A binned likelihood fit gives
$N_{D_sD_s}=3.2\pm 2.3$ signal events in the $\Delta E$ distribution, where the error 
reflects the statistical uncertainty. The corresponding value for a fit to 
the $M_{\rm bc}$ distribution is $4.7\pm 2.6$ events. 

\begin{figure}[t]
\begin{center}
\includegraphics[width=0.49\textwidth]{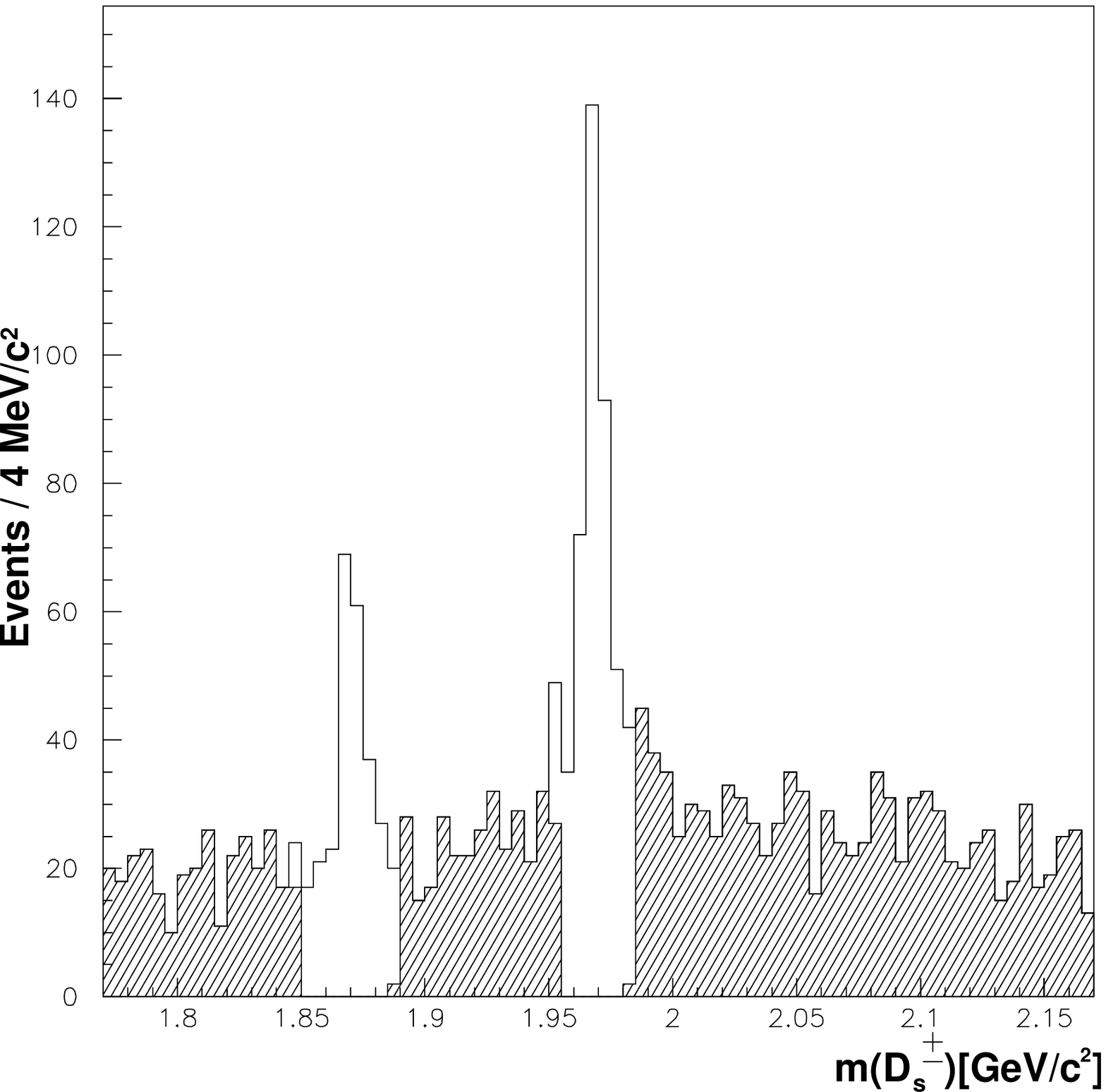}
\includegraphics[width=0.49\textwidth]{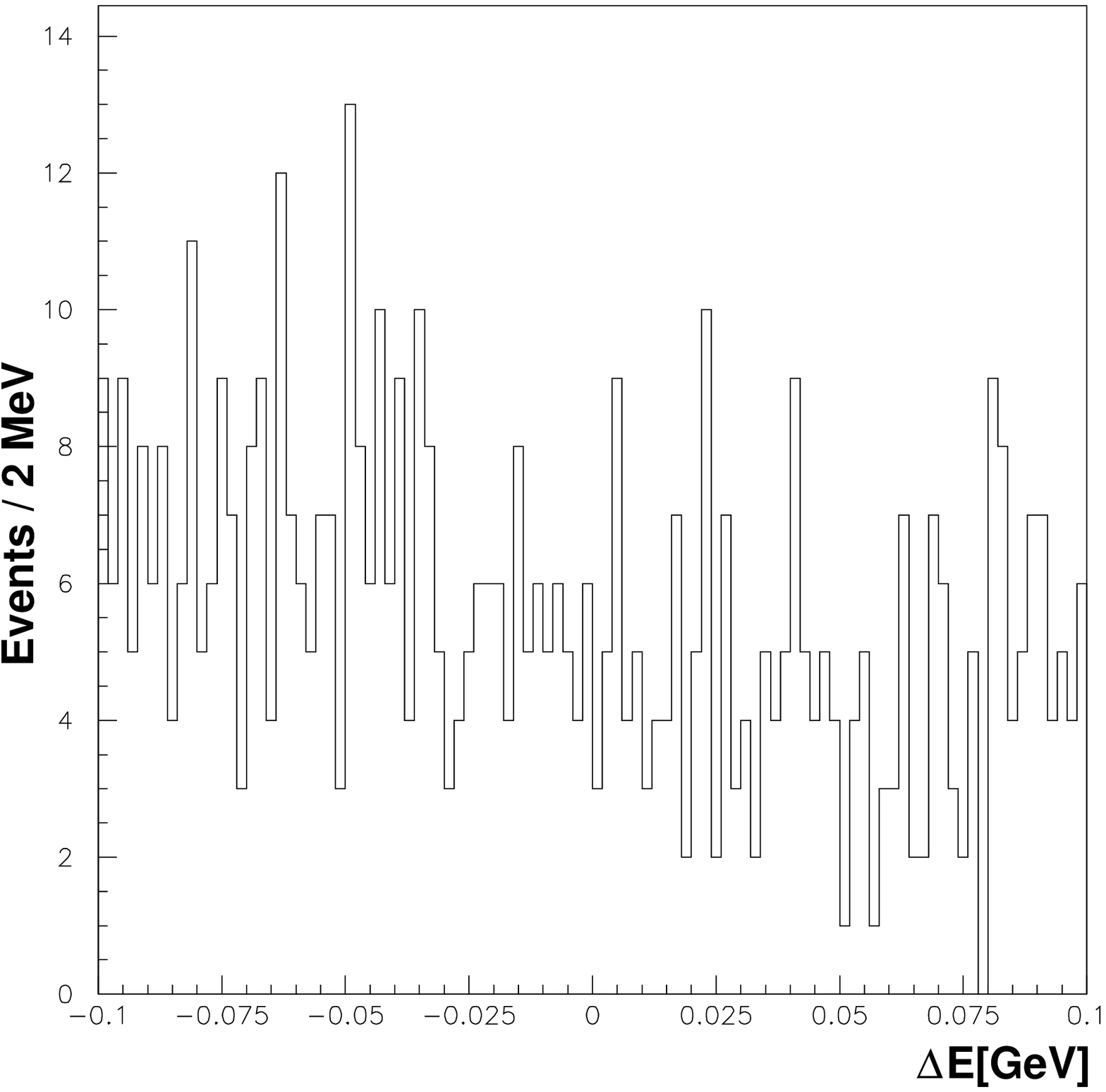}\\
\includegraphics[width=0.49\textwidth]{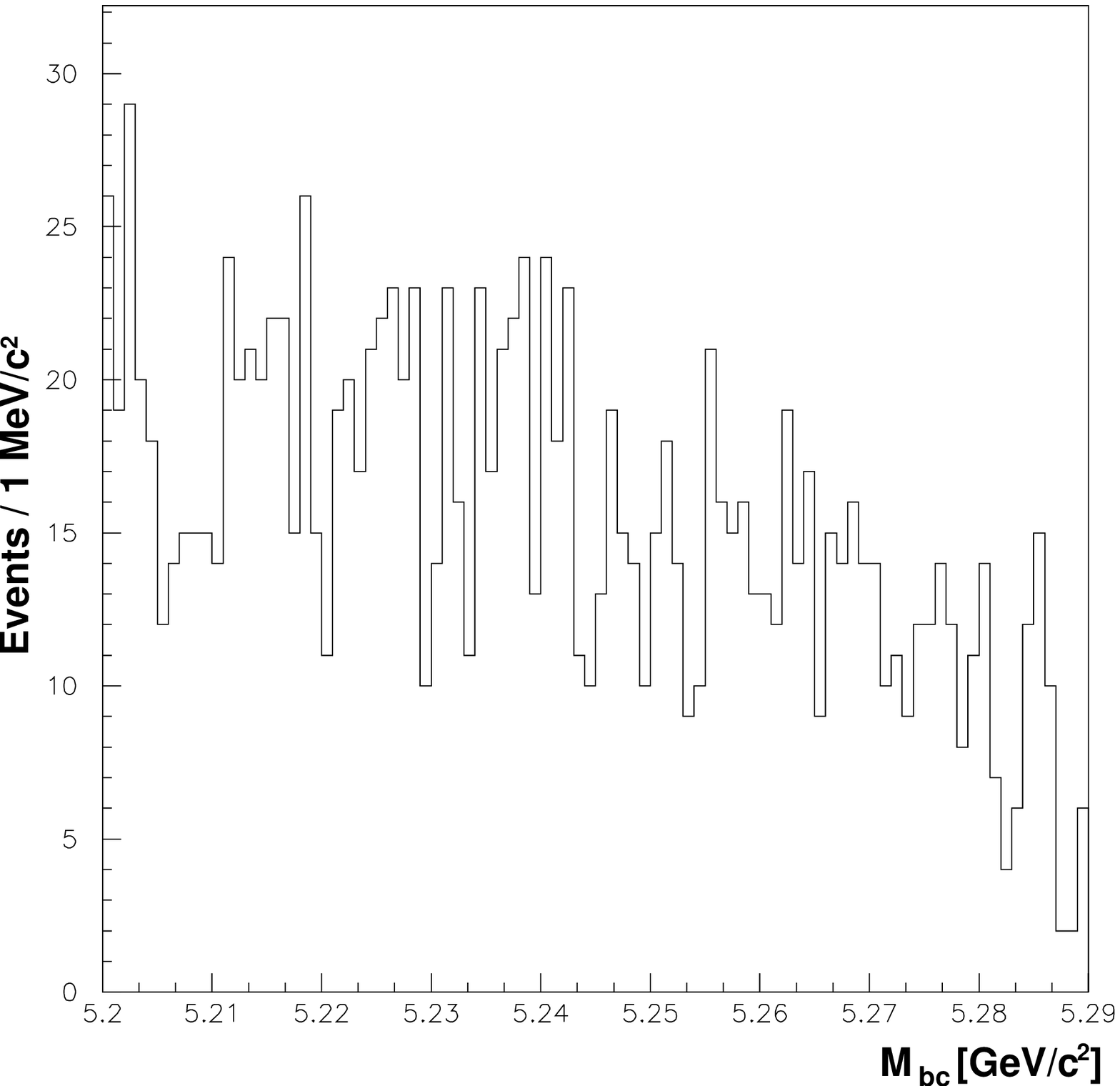}
\includegraphics[width=0.49\textwidth]{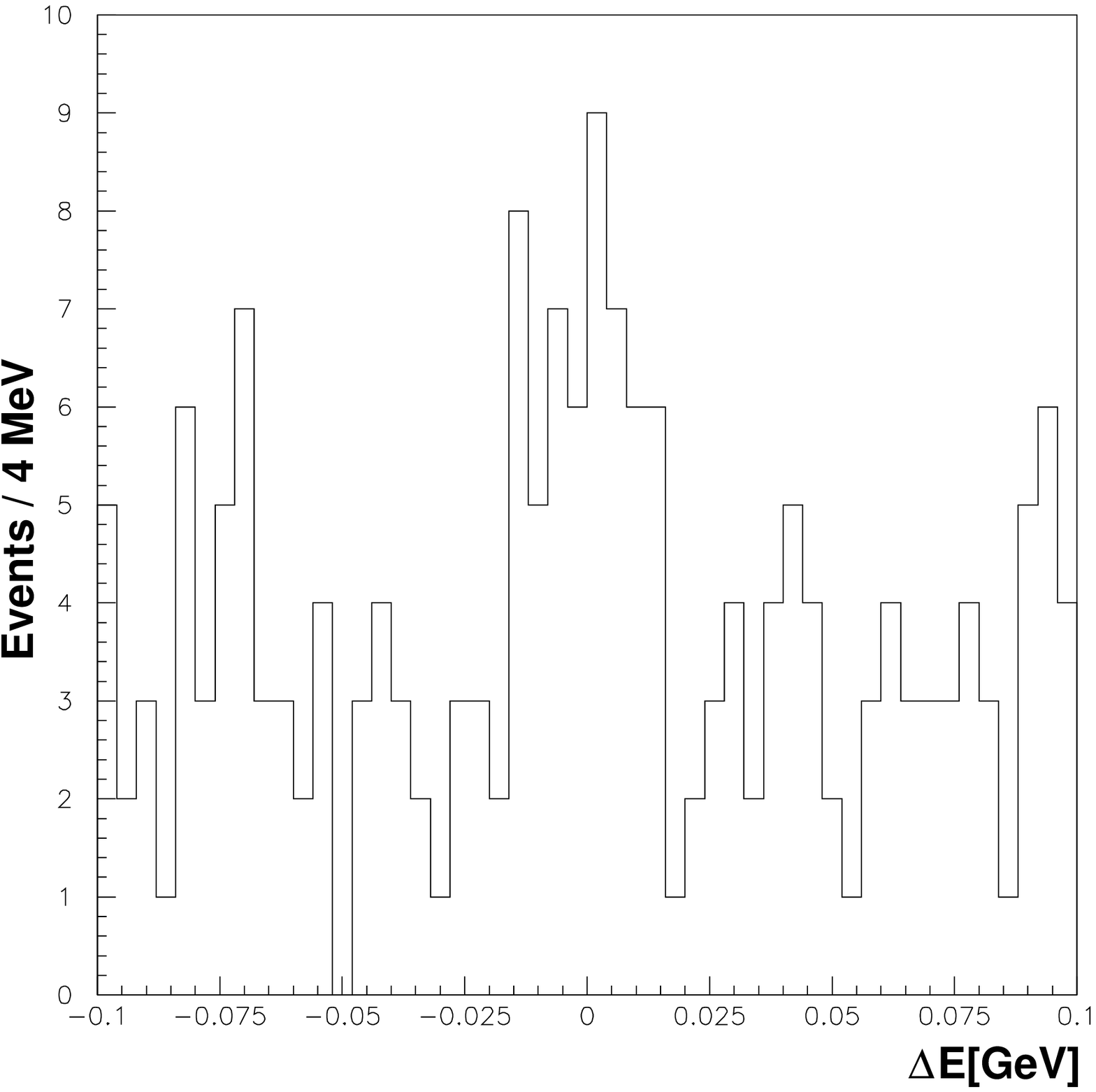}
\end{center}
\caption{Invariant mass of $D_s^\pm$ candidates with hatched sidebands (upper left).
$\Delta E$ (upper right) and $M_{\rm bc}$ (lower left) distribution for events with both $D_s$
masses in the sideband. $\Delta E$ (lower right) for $D_s^\pm$ signal candidate combined with a
$D_s^\mp$ from a sideband which includes the $D^\mp$ mass region.} 
\label{fig_13}
\end{figure}

The sidebands were used to check for a possible peaking background. 
No structure is observed in any of the $M_{\rm bc}$-$\Delta E$ distributions for
events in the $D_s$ mass sidebands (see Fig.~\ref{fig_13}). If the $D$ mass region is included 
in the $D_s$ sideband, we observe a clear contribution from $\bbar\to D_s^-D^+$ as expected.

\subsection{Results}

The number of signal $\bbar\to D_s^-D^+$ events $N_{D_sD}$ is converted into a 
branching fraction using the efficiency $\epsilon(D_sD)$ and the  number of $B\overline{B}$ events (assuming
equal production of $B^0\bbar$ and $B^+B^-$ pairs): 
${\cal B}(\bbar\to D_s^-D^+)=(7.42\pm 0.23)\times 10^{-3}$.

In addition to the above statistical error, we considered several sources of systematic
uncertainty as listed in Table~\ref{tab_4}. The largest contribution arises due to imprecise knowledge of the intermediate 
branching fractions of $D_s$ and $D$ mesons ($\pm 13.9\%$). 
We assigned a $1\%$
relative error for each of the charged tracks used in the reconstructed final states due to the
uncertainty in tracking efficiency. 
The particle identification efficiency has a relative uncertainty of $2.5\%$ per charged kaon
and $2.0\%$ per charged pion.
An additional $1.4\%$ uncertainty is due to the limited statistics of the MC
sample used to estimate the efficiency. The fraction of true signal events in the sample
($1-r_{KK\pi}-r_{DK^\ast K}$) is known to an accuracy of $\pm 2.2\%$. 
Systematics arising from
the description of the $\Delta E$ distribution are evaluated by comparing the known number of
reconstructed $\bbar\to D_s^-D^+$ events in the simulated sample with the fitted yield and 
is found to be $2.2\%$. 
Finally, the
uncertainty on the number of $B\overline{B}$ events ($1.1\%$) is taken into account. 
The sum in quadrature of the individual
contributions gives a systematic error of $18.3\%$. The measured branching fraction is thus
\begin{equation}
{\cal B}(\bbar\to D_s^-D^+)=(7.42\pm 0.23({\rm stat.})\pm 1.36({\rm syst.}))\times 10^{-3}.
\end{equation}

The same result can be expressed by separating the largest source of systematic uncertainty,
${\cal B}(D_s^-\to\phi\pi^-)~{\cal B}(\phi\to K^+K^-)$, in order to allow the result to be rescaled to future improved
measurements of this quantity. In this case, the error arising from intermediate ${\cal B}$'s reduces to
$7.7\%$~\cite{foot_3} and
${\cal B}(\bbar\to D_s^-D^+)\cdot {\cal B}(D_s^-\to\phi\pi^-){\cal B}(\phi\to K^+K^-)=
\bigl[1.47\pm 0.05({\rm stat.})\pm 0.21({\rm syst.})\bigr]\times 10^{-4}$.

We observe no statistically significant signal in the $\bbar\to D_s^+D_s^-$ decay mode. From
the fitted number of events $N_{D_sD_s}$ we derive ${\cal B}(\bbar\to D_s^+D_s^-)=(7.1\pm 5.1)\times 10^{-5}$, 
with a statistical error only. Applying the Feldman-Cousins~\cite{Feldman} upper limit
procedure to this value gives ${\cal B}(\bbar\to D_s^+D_s^-)\le 1.6\times 10^{-4}$ at 90\% C.L. 

Evaluation of systematic errors follows closely the treatment of the $\bbar\to
D_s^-D^+$ decay mode. Individual sources contributing to ${\cal B}(\bbar\to D_s^+D_s^-)$ 
are listed in Table~\ref{tab_4}. The error arising from fitting was estimated by comparing the
fitted and true number of simulated 
signal events. Additional uncertainty arises due to the use of a fixed width and position of the
Gaussian function in the fit. The fit was repeated with parameters changed
by one standard deviation; this resulted in a $4.1\%$ change in ${\cal B}$. 
The overall systematic error is found to be $28.4\%$. We inflate the upper
limit by this amount to obtain
\begin{equation}
{\cal B}(\bbar\to D_s^+D_s^-)\le 2.0\times 10^{-4}~~{\rm at}~90\% ~{\rm C.L.} 
\end{equation}

\begin{table}[t]
\caption{Sources of systematic uncertainty in ${\cal B}(\bbar\to D_s^-D^+)$ and 
${\cal B}(\bbar\to D_s^+D_s^-)$ measurements. }
\label{tab_4}
\begin{tabular}
{@{\hspace{0.5cm}}l@{\hspace{0.5cm}}||@{\hspace{0.5cm}}c@{\hspace{0.5cm}}||@{\hspace{0.5cm}}l@{\hspace{0.5cm}}||@{\hspace{0.5cm}}c@{\hspace{0.5cm}}}
\multicolumn{2}{c}{$\bbar\to D_s^-D^+$} & \multicolumn{2}{c}{$\bbar\to D_s^+D_s^-$} \\
\hline \hline
Source & Rel. syst. error $[\% ]$ & Source & Rel. syst. error $[\% ]$\\
\hline
$D_s$ and $D$ ${\cal B}$'s & $13.9$ & $D_s$ ${\cal B}$'s & $23.1$ \\ 
Tracking & $6.0$ & Tracking & $6.0$ \\ 
Kaon ident. & $7.1$ & Kaon ident. & $8.5$\\ 
Pion ident. & $6.4$ & Pion ident. & $5.2$\\ 
MC statistics & $1.4$ & MC statistics & $5.3$\\ 
$r_{KK\pi}$, $r_{DK^\ast K}$ & $2.2$ & Fixed parameters & $4.1$\\  
Fitting model & $2.2$ & Fitting model & $9.5$\\ 
$N(B\overline{B})$ & $1.1$ & $N(B\overline{B})$ & $1.1$\\ 
\hline
Total & $18.3$ & Total & $28.4$\\
\hline \hline
\end{tabular}
\end{table}

\subsection{Conclusions}
We have measured the branching fraction for $\bbar\to D_s^-D^+$ decays. The measured value is
${\cal B}(\bbar\to D_s^-D^+)=\bigl[7.42\pm 0.23({\rm stat.})\pm 1.36({\rm syst.})\bigr]\times 10^{-3}$,  
which represents a large improvement in accuracy as compared to previous measurements.

This branching fraction can be used to obtain a value
for the CKM matrix element $|V_{ub}|$ following~\cite{Kim}.
Using ${\cal B}(\bbar\to D_s^-\pi^+)$ from~\cite{Dspi}, we
obtain $|{V_{ub}/V_{cb}}|=(7.4\pm 1.2)\times 10^{-2}$. 
The error includes experimental uncertainties on the measured branching fractions 
(the largest individual contribution is due to the statistical error on ${\cal B}(\bbar\to D_s^-\pi^+)$ measurement, 
while the uncertainty on the ${\cal B}(D_s^+\to\phi\pi^+)$ cancels in the ratio), as well as the uncertainty of the theoretical prediction.
This result is in agreement with measurements from semileptonic decays~\cite{PDG}.

For $\bbar\to D_s^+D_s^-$ decays we find no statistically significant signal. We set an
upper limit of
${\cal B}(\bbar\to D_s^+D_s^-)\le 2.0\times 10^{-4}$ at 90\% C.L.
This result is the first limit on this decay mode and is below
the prediction of~\cite{Fajfer}. 

We thank the KEKB group for the excellent operation of the
accelerator, the KEK cryogenics group for the efficient
operation of the solenoid, and the KEK computer group and
the National Institute of Informatics for valuable computing
and Super-SINET network support. We acknowledge support from
the Ministry of Education, Culture, Sports, Science, and
Technology of Japan and the Japan Society for the Promotion
of Science; the Australian Research Council and the
Australian Department of Education, Science and Training;
the National Science Foundation of China under contract
No.~10175071; the Department of Science and Technology of
India; the BK21 program of the Ministry of Education of
Korea and the CHEP SRC program of the Korea Science and
Engineering Foundation; the Polish State Committee for
Scientific Research under contract No.~2P03B 01324; the
Ministry of Science and Technology of the Russian
Federation; the Ministry of Higher Education, 
Science and Technology of the Republic of Slovenia;  
the Swiss National Science Foundation; the National Science Council and
the Ministry of Education of Taiwan; and the U.S.\
Department of Energy.


\begin{thebibliography}{99}
\bibitem{foot_1} Inclusion of charge conjugate states is implicit throughout this report.
\bibitem{foot_2} Since the simulated sample does not include the $\bbar\to D^+ K^{\ast 0} K^-$ 
contribution, an exact expression for $N_{D_sD}$ would include a correction of $r_{KK\pi}\cdot r_{DK^\ast K}$, which is negligible.
\end{thebibliography}
\end{document}